\documentclass[preprint2,twocolumn,superscriptaddress]{aastex63}

\usepackage{hyperref}
\usepackage{textcomp}
\usepackage{graphicx}
\usepackage{amsmath}
\usepackage{xcolor}
\usepackage{natbib}

\newcommand{\V}[1]{\mathbf{#1}} 
\newcommand{\T}[1]{\texttt{#1}} 
\newcommand{\xhat}{\mbox{$\hat{\mathbf{x}}$}} 
\newcommand{\yhat}{\mbox{$\hat{\mathbf{y}}$}} 
\newcommand{\zhat}{\mbox{$\hat{\mathbf{z}}$}}

\newcommand{\figref}[1]{Figure~\ref{#1}}
\newcommand{\secref}[1]{\S\ref{#1}}
\newcommand{\eqnref}[1]{Equation~(\ref{#1})}

\begin{document}

\title{Identification of active magnetic reconnection using magnetic flux transport in plasma turbulence}


\correspondingauthor{Tak Chu Li}
\email{tak.chu.li@dartmouth.edu}

\author{Tak Chu Li}
\author{Yi-Hsin Liu}
\affiliation{Department of Physics and Astronomy, Dartmouth College, Hanover, NH, USA}
\author{Yi Qi}
\affiliation{Department of Earth, Planetary, and Space Sciences, University of California, Los Angeles, CA, USA}

\begin{abstract}

Magnetic reconnection has been suggested to play an important role in the dynamics and energetics of plasma turbulence by spacecraft observations, simulations and theory over the past two decades, and recently, by magnetosheath observations of MMS. A new method based on magnetic flux transport (MFT) has been developed to identify reconnection activity in turbulent plasmas. This method is applied to a gyrokinetic simulation of two-dimensional (2D) plasma turbulence. Results on the identification of three active reconnection X-points are reported. The first two X-points have developed bi-directional electron outflow jets. Beyond the category of electron-only reconnection, the third X-point does not have bi-directional electron outflow jets because the flow is modified by turbulence. In all cases, this method successfully identifies active reconnection through clear inward and outward flux transport around the X-points. This transport pattern defines reconnection and produces a new quadrupolar structure in the divergence of MFT. This method is expected to be applicable to spacecraft missions such as MMS, Parker Solar Probe, and Solar Orbiter. 
\end{abstract}




\section{Introduction}\label{sec:intro}
Magnetic reconnection and plasma turbulence are both fundamental processes ubiquitously operating throughout the universe. Reconnection has been suggested to contribute to energy dissipation \citep{Dmitruk:2004,Sundkvist:2007,Osman:2011,Osman:2012a,Markovskii:2011,Perri:2012a,Wan:2012,Karimabadi:2013,TenBarge:2013a,Wu:2013a,Zhdankin:2013,Shay:2018} and potential changes in the cascade \citep{Loureiro:2017a,Boldyrev:2017,Mallet:2017a,Franci:2017,Mallet:2017b,Loureiro:2017b,Vech:2018a,Stawarz:2019} of turbulence by {\it in situ} observations, numerical simulations and theory. In heliospheric turbulence, reconnection was first observed {\it in situ} in the terrestrial magnetosheath by Cluster \citep{Retino:2007}. Recently, high resolution measurements from MMS \citep{burch16a} have enabled the detection of electron jets in small-scale current sheets in the turbulent magnetosheath \citep{Yordanova:2016,Voros:2017,Phan:18,Wilder:2018}, including notably, electron-only reconnection \citep{Phan:18}.


Reconnection occurs in a small-scale electron diffusion region (EDR) within a thin current sheet. As upstream field lines flow into the EDR, they reconnect at the X-point. The reconnected field possesses strong magnetic tension, which drives the reconnected field away from the X-point, ejecting plasma that is coupled to it as bi-directional outflow jets. The fundamental process of reconnection can be described as inward and outward transport of magnetic flux and associated plasmas at an X-point. The transport of magnetic flux and plasma flows across a separatrix was used to to define reconnection \citep{Vasyliunas:1975}.

At the frontier of turbulence and reconnection research, important questions include how reconnection occurs in a dynamical turbulent system and how the rich dynamics of turbulence and reconnection, such as turbulent energy dissipation and cascade, interplay. Nevertheless, there is still no clear, reliable method to identify reconnection X-points in turbulent plasmas. 
In 2D turbulence simulations, the method of saddle points that define an X-point topology was applied \citep{servidio09a,Servidio:2010,Wan:2013,Haggerty:2017}. However, among a large number of identified X-points, only a few displayed significant reconnection electric fields \citep{servidio09a}. It would be possible that many identified X-points are not actively reconnecting.


In observations, a commonly used method to identify reconnection is the detection of bi-directional Alfv\'enic ion outflow jets. In a turbulent system such as the terrestrial magnetosheath, reconnection can happen at sub-ion or electron scales \citep{Wilder:2018,Phan:18}, and electron jets becomes the conclusive signature of reconnection. However, fast turbulent flows at sub-ion scales can make the detection challenging. In fact, only one out of several tens of sub-ion-scale current sheets detected by \citet{Phan:18} displayed clear bi-directional reconnection electron jets. 

Recently, the transport of magnetic flux around an X-point was considered in kinetic simulations of reconnection \citep{YHLiu:16,YHLiu:18c}. MFT takes into account the decoupling of electron flow and magnetic flux (slippage) arising from a non-ideal electric field, and thus correctly captures the inward and outward transport of magnetic flux around a reconnection X-point. In a symmetric reconnection simulation with shear flows, the electron flow can be highly distorted \citep{YHLiu:18c}. Under stronger shear flows or asymmetry likely in turbulence, the electron flow may not show typical reconnection outflows. In fact, in a highly asymmetric configuration, active reconnection with only one electron jet is possible \citep{YHLiu:16}.



\section{Theory}\label{sec:theory}
The transport of magnetic flux inherent to reconnection represents an innovative way for identifying active reconnecting X-points in turbulence. The presence of inward flux transport also indicates reconnection is actively taking place. The MFT velocity $\mathbf{U}_\psi$ was previously derived in one and two dimensions \citep{YHLiu:16,YHLiu:18c}. The key steps leading to the definition of $\mathbf{U}_\psi$ are summarized here. In 2D, the magnetic field can be represented as an in-plane and out-of-plane (guide field) component directed along $\zhat$: $\mathbf{B} = \zhat\times\nabla\psi + B_0\zhat$. Curling the Faraday's law: $\zhat\times[\, \partial_t\mathbf{B} + c\nabla\times\mathbf{E} = 0 \,]$ results in $\partial_t\psi = c E_z$. We then consider the electron momentum equation:
$\mathbf{E}+ \mathbf{v}_e\times\mathbf{B}/c = \mathbf{E}_e'$, where $\mathbf{E}_e'$ is the non-ideal electric field in the electron frame. 
Taking the $z$ component of this equation and casting it into 
the form of the 2D advection equation of magnetic flux: $\partial_t\psi + \mathbf{U_\psi}\cdot\nabla_\perp \psi = 0$, the in-plane MFT velocity is then given by:
\begin{eqnarray}
\mathbf{U}_\psi\equiv \mathbf{v}_{ep} - (\mathbf{v}_{ep}\cdot\hat{b}_p)\hat{b}_p + \frac{cE_{ez}'}{B_p}(\zhat\times\hat{b}_p),
\label{eq:Upsi}
\end{eqnarray}
where $\hat{b}_p \equiv \mathbf{B}_p/B_p$ is the unit vector of the in-plane magnetic field $\mathbf{B}_p$ and $\mathbf{v}_{ep}$ the in-plane electron flow. The first two terms represent the in-plane electron flow perpendicular to $\mathbf{B}_p$. They come from the $\mathbf{v_e}\times\mathbf{B}$ term in the electron momentum equation. For $\mathbf{E}_e'$=0, the electron flow is frozen-in to the magnetic field and they move together. When $\mathbf{E}_e'$$\not=$0, slippage between magnetic flux and electron flow arises as the last term. Without separating the perpendicular electron flow and slippage terms, which provide a relation between the transport of magnetic flux and electron flow, \eqnref{eq:Upsi} can be simplified to:
\begin{eqnarray}
 \mathbf{U}_\psi = \frac{cE_{z}}{B_p}(\zhat\times\hat{b}_p).
\label{eq:Upsi2}
\end{eqnarray}To the first order in gyrokinetics, $\mathbf{U}_\psi$ is given by \eqnref{eq:Upsi} or (2) with $\mathbf{v}_{ep}$, $\mathbf{B}_p$ and $E_{ez}'$ replaced by $\delta\mathbf{u}_{ep}$, $\delta \mathbf{B}_p$ and $\delta E_{ez}'= \delta E_z + (\delta\mathbf{u}_{ep} \times \delta\mathbf{B}_{p}/c)_z$,
where fluctuating quantities in turbulence are the in-plane electron bulk flow $\delta\mathbf{u}_{ep}$ and so on. Note that \eqnref{eq:Upsi} is not applicable at the X-point because a source or sink term, representing flux generation or annihilation at the X-point, is not included in this advection equation.

A new quantity, the divergence of MFT, $\nabla\cdot \mathbf{U}_\psi$, is considered here. $\nabla\cdot \mathbf{U}_\psi <0$ and $>0$ can capture the converging inflows and diverging outflows of magnetic flux, respectively. These bi-directional inflows and outflows of magnetic flux at an X-point signifies active reconnection. 
$\nabla\cdot \mathbf{U}_\psi$ also informs about the time scale of diverging magnetic flux from the X-point. 
Having the dimension of inverse time, $\nabla\cdot \mathbf{U}_\psi$ is frame-independent in 2D in the non-relativistic limit. Therefore, one can compute $\nabla\cdot \mathbf{U}_\psi$ for moving X-points without changing frames. 


\section{Code}
The 2D gyrokinetic turbulence simulation has been previously performed \citep{Li:2016} using the the Astrophysical Gyrokinetics Code, or \T{AstroGK}, described in details in \citep{Numata:2010}. \T{AstroGK} has been extensively used to investigate turbulence in weakly collisional plasmas \citep{Howes:2008a,Tatsuno:2009,Howes:2011a,TenBarge:2012a,Nielson:2013a,TenBarge:2013a,Howes:2016b,Li:2016,Howes:2018a,Li:2019}
and collisionless strong-guide-field reconnection \citep{Numata:2011,TenBarge:2014b,Kobayashi:2014,Numata:2015}.
\T{AstroGK} is an Eulerian continuum code with triply periodic
boundary conditions. It has a slab geometry elongated along the
straight, uniform background magnetic field, $\V{B}_0=B_0 \zhat$. The
code evolves the perturbed gyroaveraged Vlasov-Maxwell equations in
five-dimensional phase space (three-dimensional-two-velocity)
\citep{Frieman:1982,Howes:2006}. The evolved quantities are the
electromagnetic gyroaveraged complementary distribution function
for each species $s$, the scalar
potential $\varphi$, parallel vector potential $A_\parallel$ and
parallel magnetic field perturbation $\delta B_\parallel$, where
$\parallel$ is along the total local magnetic field $\V{B}=B_0\zhat+
\delta \V{B}$. The total and background magnetic fields are the same 
to first-order accuracy retained for perturbed fields in gyrokinetics.
The velocity grid is specified by pitch angle $\lambda=v_\perp^2/v^2$ and 
energy $\varepsilon=v^2/2$. 
The background distribution functions for both species are stationary
uniform Maxwellians. Collisions are incorporated using a fully
conservative, linearized gyro-averaged Landau collision operator \citep{Abel:2008,Barnes:2009}.

\section{Setup}
The 2D Orszag-Tang Vortex (OTV) problem has been widely
used to study plasma turbulence \citep{Politano:1989,Dahlburg:1989,Picone:1991,Politano:1995b,Grauer:2000,Mininni:2006,Parashar:2009,Parashar:2014b}. It is given by
\begin{equation}
 \begin{aligned}
  \delta\mathbf{u} &= \delta u [ - \sin(k_\perp y) \xhat + \sin(k_\perp x)\yhat]\nonumber\\
  \delta\mathbf{B} &= \delta B [ - \sin(k_\perp y)\xhat +  \sin(2k_\perp x) \yhat],\\
 \end{aligned}
\label{eq:vB2D}
\end{equation}
where $\delta u =\delta B/\sqrt{4 \pi \rho_0}$, $\delta \V{u}$ and $\delta \V{B}$ are perturbations in the ion and electron bulk flow and the magnetic field, and $k_\perp=2\pi/L_\perp$ are positive constants.

To follow the turbulent cascade from the inertial range ($k_\perp\rho_i \ll 1$) to below electron scales ($k_\perp\rho_e >1$) \citep{TenBarge:2013a,TenBarge:2013b,TenBarge:2014b}, we specify a reduced mass ratio, $m_i/m_e=25$, which, in a simulation domain of $L_{\perp}=8\pi \rho_i$ and dimensions $(n_x,n_y,n_z,n_\lambda,n_\varepsilon,n_s)=(128,128,2,64,32,2)$, enables us to resolve a dynamic
range of $0.25 \le k_\perp\rho_i \le 10.5$, or $0.05 \le k_\perp\rho_e \le 2.1$. 
Plasma parameters are ion plasma $\beta_i = 8 \pi n_i T_{0i}/B_0^2=0.01$ and $T_{0i}/T_{0e}=1$. Collision frequencies of $\nu_i$ =
10$^{-5} \omega_{A0}$ and $\nu_e$ = 0.05 $\omega_{A0}$ (where
$\omega_{A0} \equiv k_{\parallel}v_A$ is a characteristic Alfv\'en wave
frequency in 3D) are sufficient to keep velocity space well resolved
\citep{Howes:2008a,Howes:2011a}.
Length, time and velocity are normalized to the ion gyroradius $\rho_i\equiv v_{ti}/\Omega_{ci}$, where $\Omega_{ci}\equiv eB_0/m_ic$, domain turnaround time $\tau_0\equiv L_\perp/\delta u$ and electron thermal speed $v_{te}\equiv\sqrt{2T_{0e}/m_e}$. $\tau_0$ can be converted to the inverse ion gyro-frequency, a relevant time scale for reconnection, by $\tau_0$ = 25 $\Omega_{ci}^{-1}$.
The divergence of velocity is normalized to $v_{te}/\rho_e=\Omega_{ce}$.


\begin{figure}[t!]
\resizebox{3.5in}{!}{\includegraphics[scale=1.,trim=1.4cm 0.1cm 1.4cm 0.1cm, clip=true]{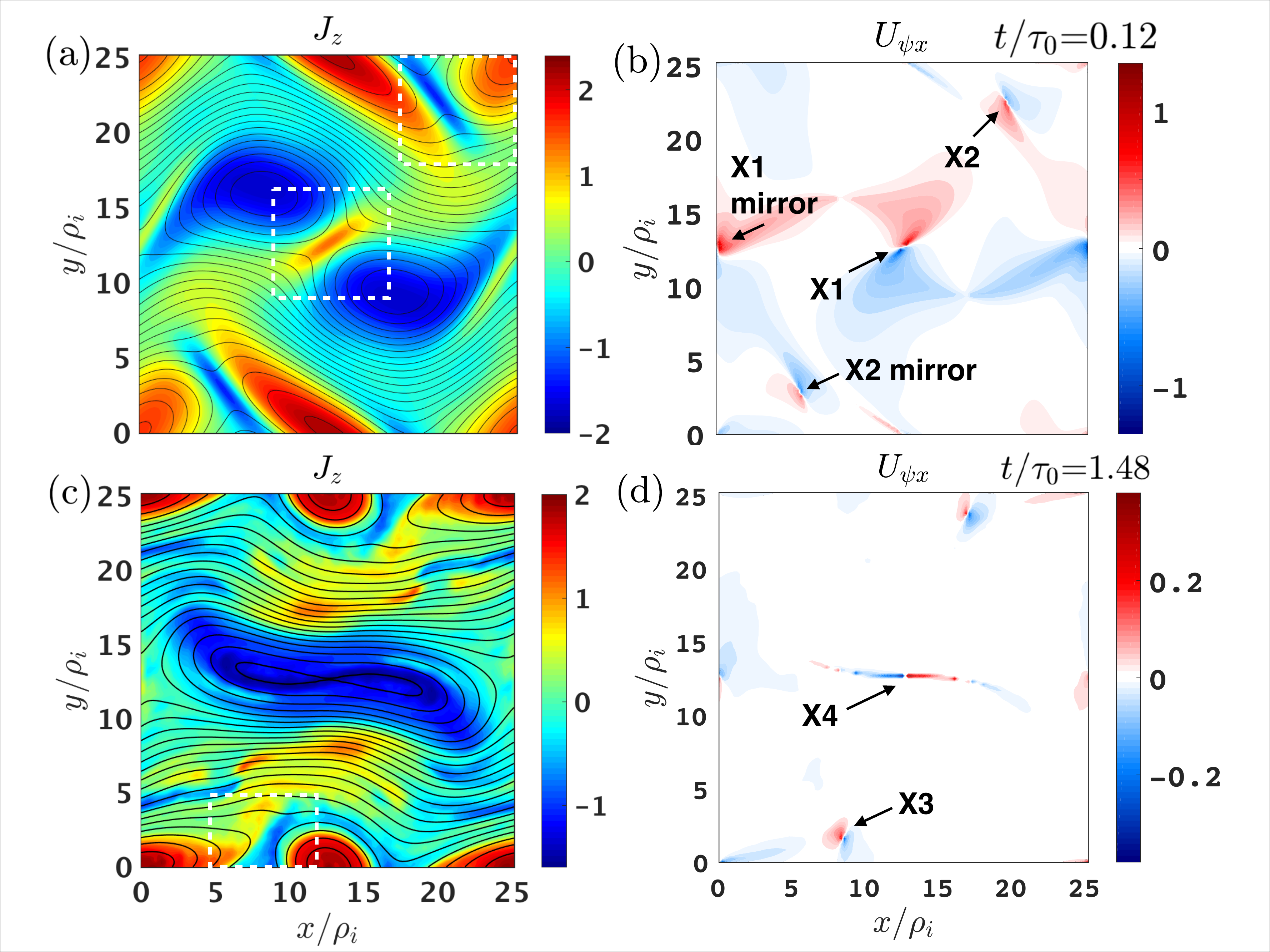}}
\caption{ \label{fig:f1} (a) The out-of-plane current density $J_z$ (color) overlaid with contours of $A_\parallel$ of the OTV configuration, and (b) the $x$-component of $\mathbf{U}_\psi$ at $t/\tau_0$=0.12, showing X1 and X2 and their mirrors (labeled). At $t/\tau_0$=1.48, (c) $J_z$ showing developed turbulence and (d) $U_{\psi x}$ revealing X3, X4 (labeled) and X3 mirror. The bi-directional outflows of magnetic flux at X1 and X4, and inflows at X2 and X3 are observed. $\delta B_p$ is offsetted by adding a 1\% of its maximum value in the domain such that $\mathbf{U}_\psi$ remains finite at the X- and O-points that have vanishing $\delta B_p$. This does not qualitatively affect the profile of $\mathbf{U}_\psi$. Dashed boxes indicate regions in zoomed-in figures.
See an animation of $J_z$ online. The animation lasts for $t/\tau_0$=0--2.01. It shows the evolution of reconnection driven by turbulent flows.
}
\end{figure}

\section{Results}

\figref{fig:f1}(a) shows the out-of-plane current density $J_z$ (color) and contours of the parallel vector potential $A_\parallel$ representing magnetic field lines of the OTV at an early time of $t/\tau_0$=0.12. The OTV has an initial flow configuration that rotates the two vortices near the center of the domain, forming a current sheet in between. The symmetry of the two vortices allows symmetric reconnection to take place at the current sheet. The flows also drive two asymmetric vortices at the top right and bottom left, resulting in two mirroring asymmetric reconnection X-points by symmetry of the system. A fourth reconnection X-point, which is a mirror of the central symmetric reconnection X-point, is located at $(x,y)\simeq$(0,12.6). The central symmetric (X1) and top-right asymmetric (X2) X-points are two of the cases we will discuss in details.


As the total turbulence energy dissipates over time \citep{Li:2016}, the driving of reconnection weakens and reconnection at later times is generally weaker than early-time events. \figref{fig:f1}(c) shows $J_z$ at late time $t/\tau_0$=1.48 when multiscale features, including small-scale current sheets, have developed. A turbulent cascade in the dissipation range (see \figref{fig:fA1} for the magnetic energy spectrum) is also developed.
At this time, an asymmetric reconnection X-point forms at the bottom left. This X-point (X3) does not develop bi-directional electron outflow jets and therefore cannot be identified through electron flows. Below we discuss the application of MFT and the identification of each reconnection X-point.

\figref{fig:f1}(b) shows the $x$-component of the MFT velocity, $U_{\psi x}$, of the whole domain at $t/\tau_0$=0.12, showing X1 and X2 as well as their mirrors, and (d) at $t/\tau_0$=1.48, showing X3, its mirror and a reconnection X-point (X4) formed at the center of an evolved, elongated vortex (flux tube). The factor of $\delta B_p^{-1}$ in the definition of $\mathbf{U}_\psi$ could tend to infinity at the X- and O-points where $\delta B_p$ vanishes.
As a practical step, we add a 1\% offset to $\delta B_p$ everywhere so that $\mathbf{U}_\psi$ remains finite at the X- and O-points that have vanishing $\delta B_p$. For the range of 0.01--4\% offsets, the amplitudes of $\mathbf{U}_\psi$ and $\nabla\cdot \mathbf{U}_\psi$ only vary by a factor of 2. Note that masking the X-points by a grid point of size $\rho_e$ yields similar amplitudes to applying a 1\% $\delta B_p$ offset.
Below we zoom in to X1--X3 to investigate the X-points more thoroughly.


\begin{figure}[t!]
\resizebox{3.5in}{!}{\includegraphics[scale=1.,trim=0.2cm 0.1cm 1.6cm 0.1cm, clip=true]{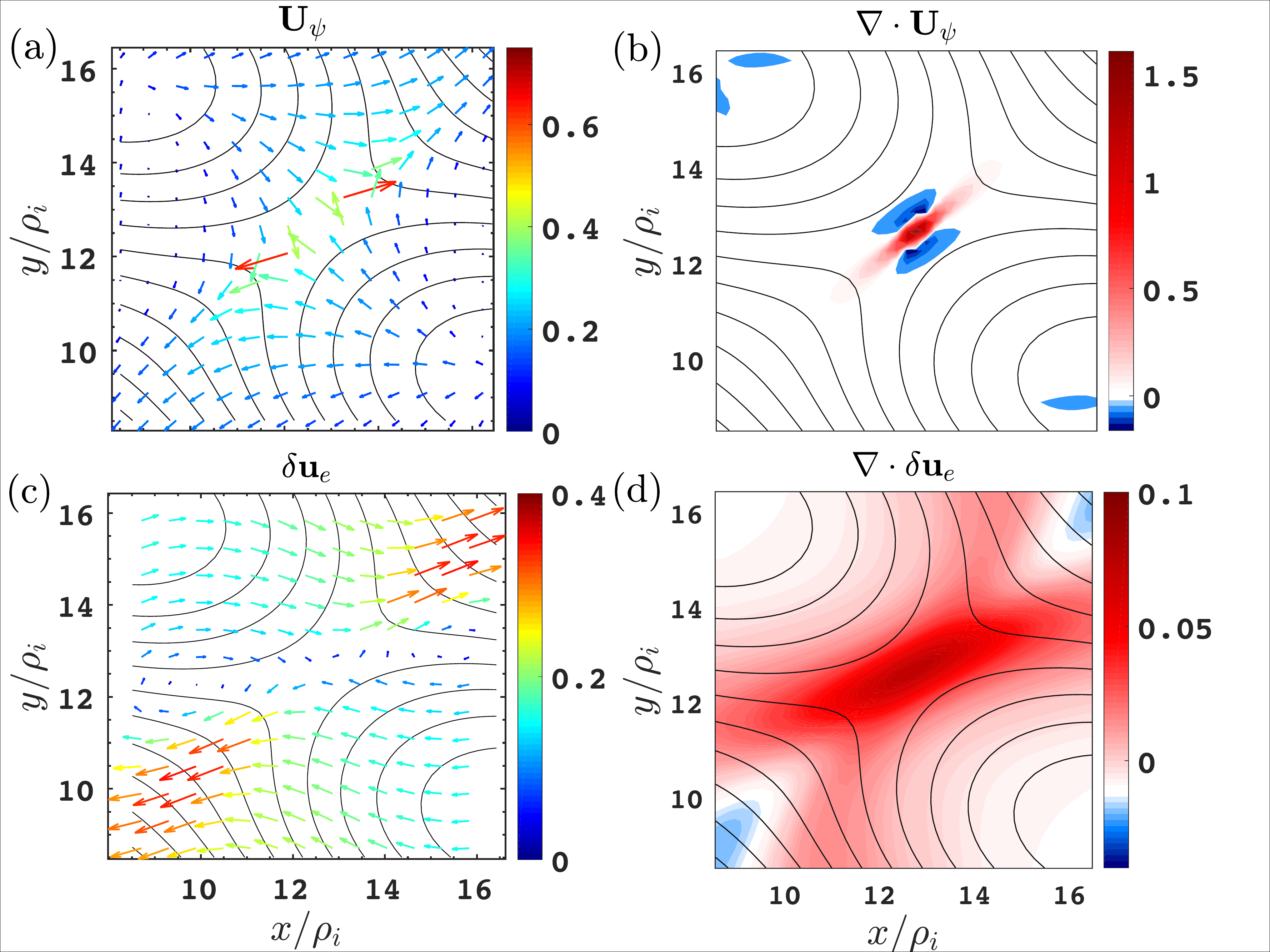}}
\caption{ \label{fig:f2} Application of the MFT method to X1, a symmetric reconnection X-point. Plotted quantities are (a) vectors of $\mathbf{U}_\psi$, (b) the divergence of $\mathbf{U}_\psi$, (c) vectors of the fluctuating in-plane electron flow velocity $\delta \mathbf{u}_e$, and (d) the divergence of $\delta\mathbf{u}_e$, overlaid with $A_\parallel$ contours. The amplitudes of vectors are denoted by the color and relative length of the arrows. The divergence of velocity is normalized to $v_{te}/\rho_e$. } 
\end{figure}

\begin{figure}[th]
\resizebox{3.5in}{!}{\includegraphics[scale=1.,trim=1.5cm 0.1cm 0.7cm 0.1cm, clip=true]{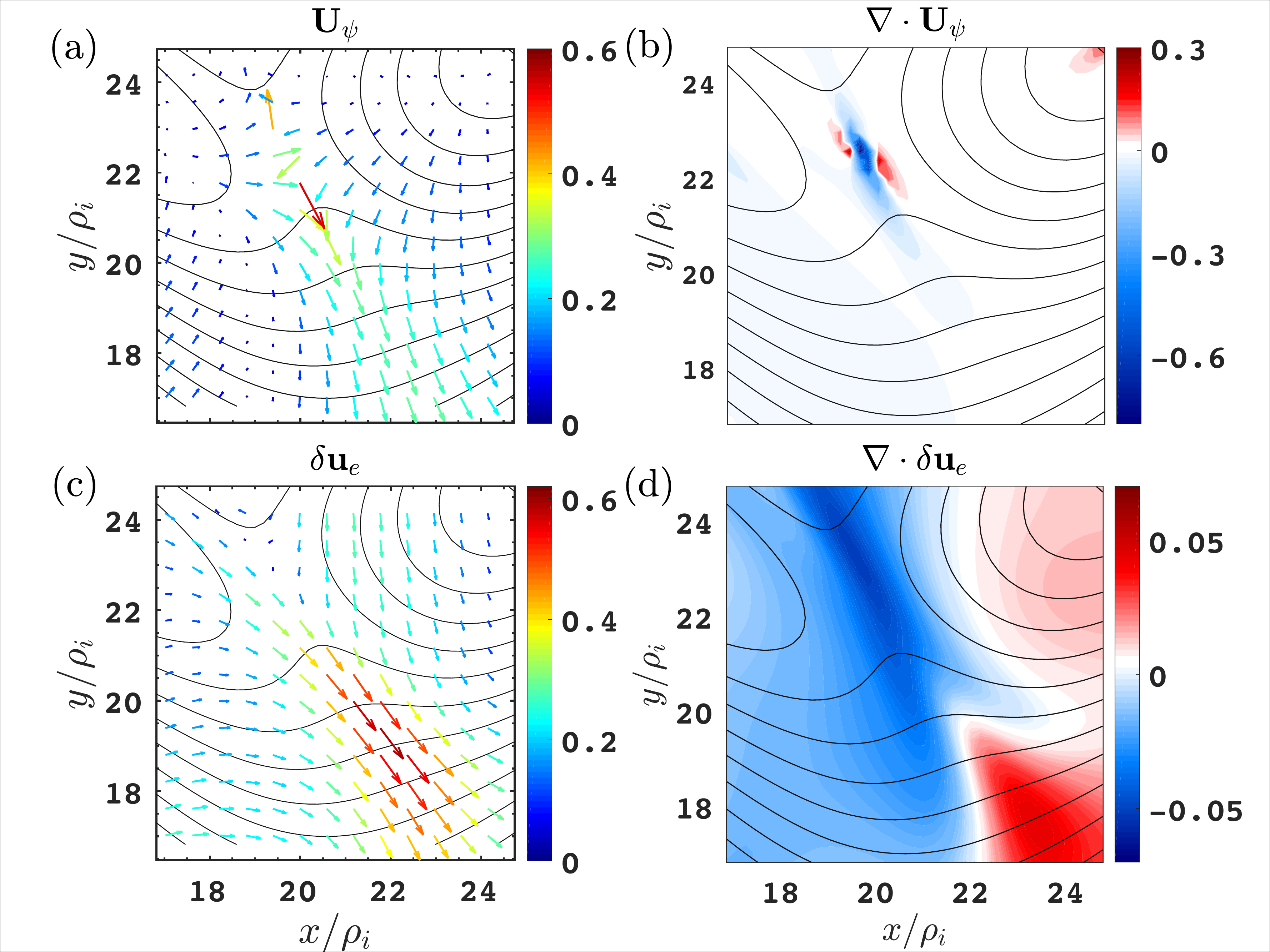}}
\caption{ \label{fig:f3} Same quantities as \figref{fig:f2} plotted for X2, an asymmetric reconnection X-point. }
\end{figure}


\subsection{X1: Symmetric reconnection X-point}
\figref{fig:f2} shows (a) vectors of $\mathbf{U}_\psi$, (b) $\nabla\cdot \mathbf{U}_\psi$, and for comparison, (c) vectors and (d) the divergence of $\delta\mathbf{u_e}$ in a zoomed-in region around X1. Clear bi-directional outflows and converging inflows of magnetic flux around X1 are captured in $\mathbf{U}_{\psi}$. (b) $\nabla\cdot \mathbf{U}_\psi$ reveals negative (blue) and positive (red) amplitudes highly localized to X1, representing converging inward and diverging outward MFT at the X-point. This is the inherent flux transport pattern of reconnection. It results in a new quadrupolar structure in $\nabla\cdot \mathbf{U}_\psi$. The quadrupolar structure reflects the bi-directional flux transport at the two sides upstream and downstream of the X-point. Both quantities are highly localized to the X-point, and can serve as local signatures of reconnection.

Bi-directional electron outflow jets in the outflow region can be seen in (c) $\delta\mathbf{u_e}$. (d) $\nabla\cdot \delta\mathbf{u}_e$ reveals positive amplitude, representing the diverging outflows. In comparison to $\mathbf{U}_\psi$, the electron outflow develops further from the X-point and over a much broader region. 


\subsection{X2: Asymmetric reconnection X-point}
The same quantities as \figref{fig:f2} are plotted around X2 in \figref{fig:f3}. Similarly, clear bi-directional inflows and asymmetric bi-directional outflows of magnetic flux are captured in (a) $\mathbf{U}_\psi$, with the downward transport being stronger. (b) $\nabla\cdot\mathbf{U}_\psi$ reveals the presence of converging inward and diverging outward flux transport as $\nabla\cdot\mathbf{U}_\psi<0$ and $>0$, respectively, at X2. Both signify active reconnection. 


In (c) $\delta\mathbf{u_e}$, asymmetric electron outflow jets are seen, with a stronger downward jet from X2. (d) The divergence of the electron flow reveals negative and positive amplitudes located broadly around and downstream from the X-point, representing converging inflows and diverging outflows of electrons at this X-point.


\begin{figure}[th]
\resizebox{3.5in}{!}{\includegraphics[scale=1.,trim=0.5cm 4.9cm 0.1cm 2cm, clip=true]{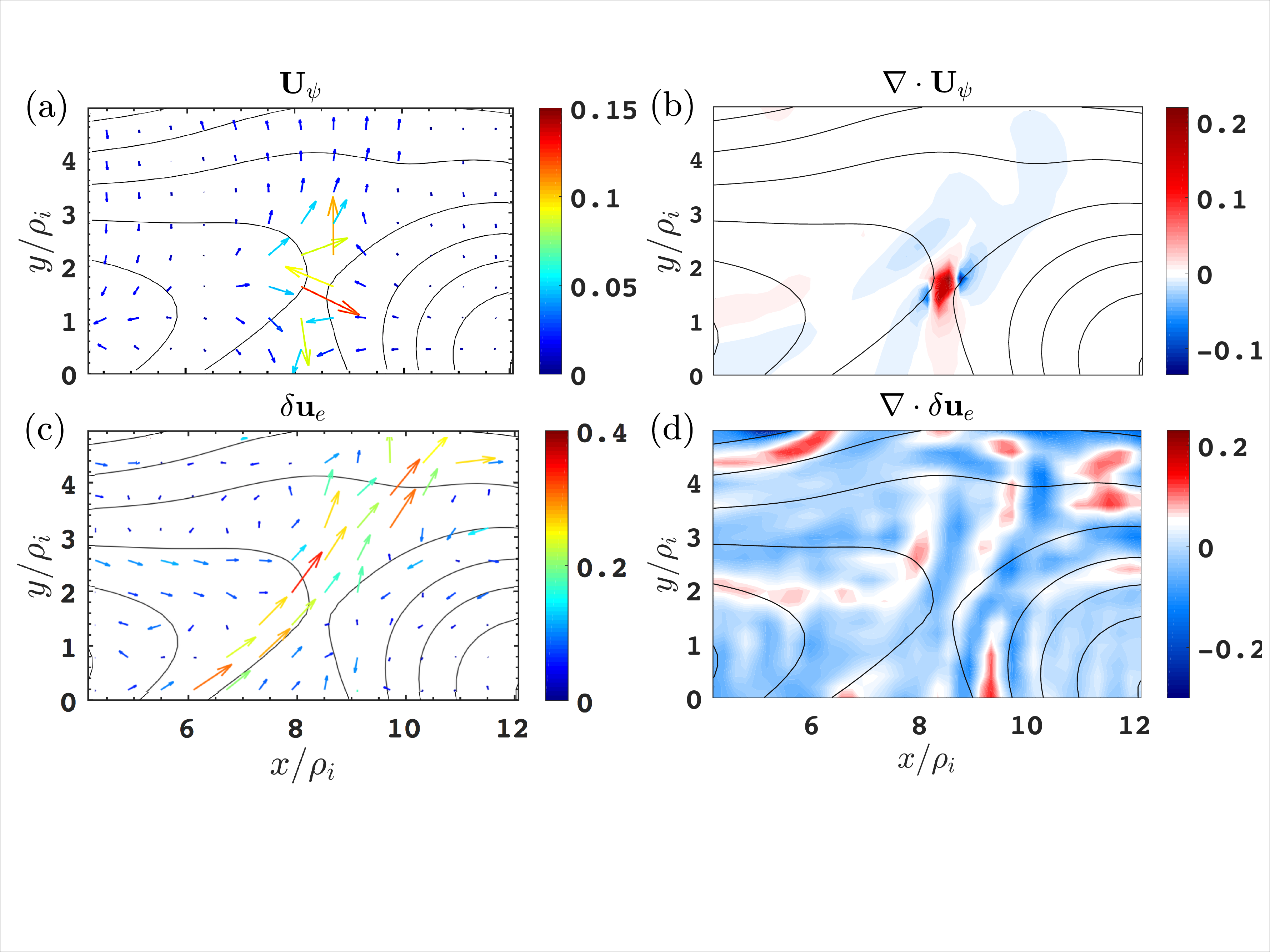}}
\caption{ \label{fig:f4} Same format as \figref{fig:f2} plotted for X3, a reconnection X-point with only one outflow jet in electrons (and ions), at $t/\tau_0 = 1.48$. Velocity vectors are measured at the X-point frame. } 
\end{figure}



\subsection{X3: Reconnection X-point Without bi-directional plasma jets} 
As the turbulent flows that drive reconnection are significantly dissipated at late times \citep{Li:2016}, reconnection activity becomes weaker than early-time reconnection. Nevertheless, converging inflows and bi-directional outflows of magnetic flux are captured in \figref{fig:f4}(a) $\mathbf{U}_\psi$ at X3. (b) $\nabla\cdot \mathbf{U}_\psi$ also reveals positive and negative amplitudes highly localized to the X-point, sharing a similar quadrupolar structure to X2 and X1.

On the other hand, the electron flow is highly modified by turbulence. No clear bi-directional outflow jets are seen in the electrons (or in the ions (Appendix \figref{fig:fA2})) at X3. Only one electron and one ion outflow jet are present. (d) $\nabla\cdot\delta\mathbf{u}_e$ also does not show clear evidence of reconnection. Plasma flows cannot be used for identifying reconnection at this X-point. However, the MFT method is able to identify reconnection through its clear inward and outward flux transport at this X-point, demonstrating the sensitivity of MFT in identifying reconnection activity in turbulence.

\subsection{Super-Alfv\'enic $\mathbf{U}_\psi$}
While $\mathbf{U}_\psi$ is normalized to $v_{te}$, it is meaningful to compare it with the upstream Alfv\'en speed. Using the electron plasma $\beta_e\equiv (v_{te}/c_{Ae})^2$=0.01, where $c_{Ae}=\sqrt{B_0/4\pi n_0m_e}$, in the simulation, and estimates of the upstream $\delta B_p/B_0\sim$ 0.1 and density $n/n_0\sim$ 0.7--1.1 for the three X-points, we can relate the upstream electron Alfv\'en speed \citep{cassak07b} to $v_{te}$ as $c_{Ae,p}/v_{te}\sim$ 1. Therefore, at X1 and X2, $U_\psi$ is of order $c_{Ae,p}$. The flux transport velocity is electron Alfv\'enic. Similarly, at X3, $U_\psi\sim$ 1.2 $c_{A,p}$ is super-Alfv\'enic. The higher velocity at early-time reconnection is associated with strong driving by initial turbulent flows. The Alfv\'enic velocity at late times is consistent with undriven reconnection simulations \citep{YHLiu:16}. $\mathbf{U}_\psi$ is between orders $c_{A,p}$ and $c_{Ae,p}$ based on the simulation.



\begin{figure}[t]
\resizebox{3.5in}{!}{\includegraphics[scale=1.,trim=0.2cm 0.1cm 1.cm 0.3cm, clip=true]{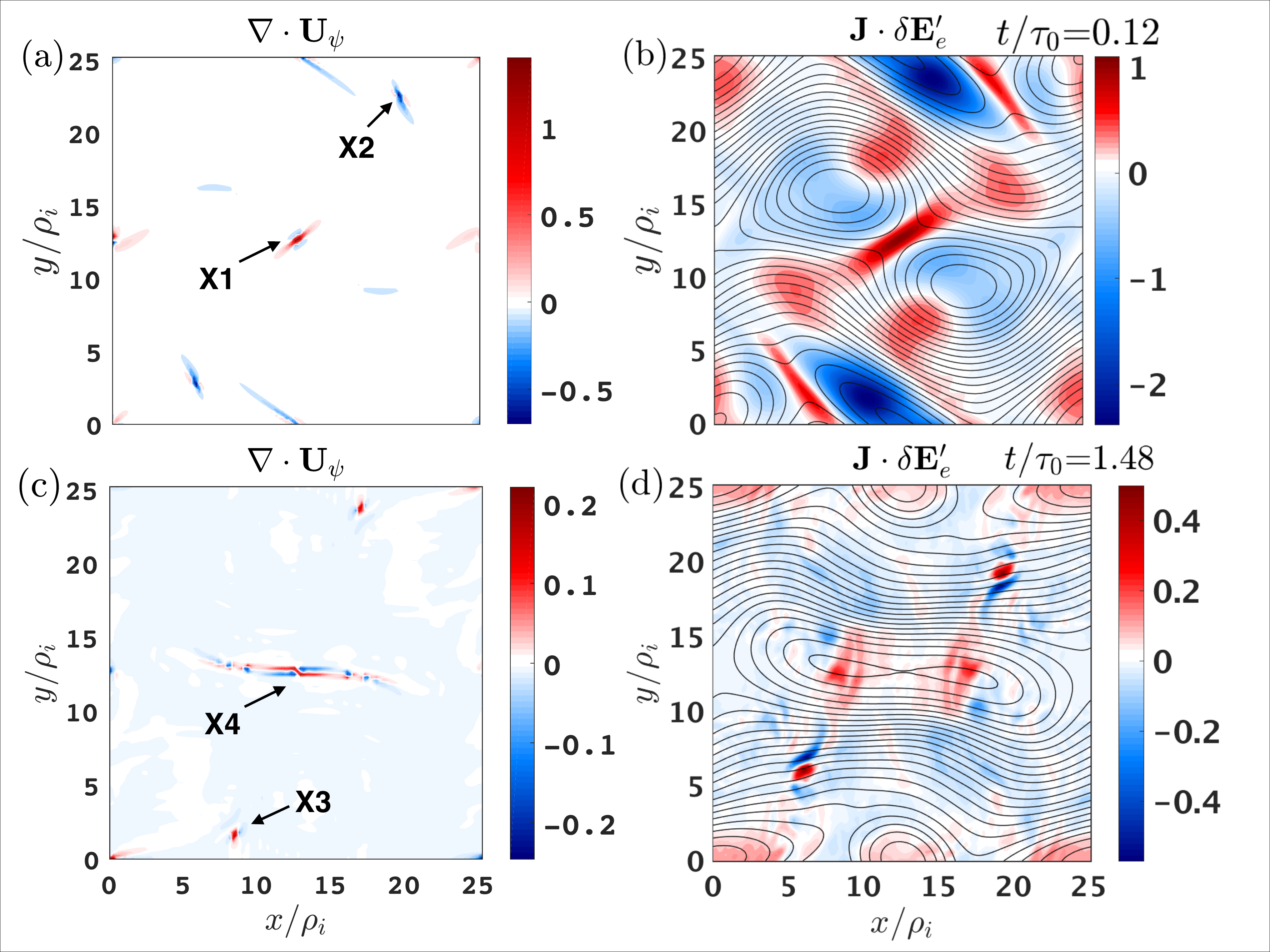}}
\caption{ \label{fig:f5} (a) The divergence of MFT of the whole domain at (a) $t/\tau_0$=0.12, showing X1 and X2 (labeled) and their mirrors, and (c) $t/\tau_0$=1.48, showing X3 and X4 (labeled) among the turbulence; (b) and (d) $\mathbf{J}\cdot\delta \mathbf{E}_e'$, the non-ideal energy conversion in the electron frame, at the two times.}

\end{figure}
\subsection{Divergence of MFT}


Plotted in \figref{fig:f5} is the divergence of MFT of the whole domain at (a) $t/\tau_0$=0.12, showing X1 and X2 and their mirrors, and at (c) $t/\tau_0$=1.48, when turbulence is developed, revealing X3 and X4. $\nabla\cdot\mathbf{U}_\psi$ shows significant amplitudes only at the active reconnection X-points, even among the turbulence. It remains small throughout the domain, and is thus suitable for the identification of reconnecting X-points in turbulence. 
$\nabla\cdot\delta\mathbf{u}_e$ is much more structured throughout the system, and at late times, becomes highly turbulent (not shown).
For comparison, (b) $\mathbf{J}\cdot\delta \mathbf{E}_e'$, energy conversion \citep{zenitani11c} in the electron frame, is much more broadly distributed over the current sheets and throughout the system. (d) At late times, it is dominated by turbulent flows far away from the reconnection X-points, and thus may not help in locating reconnection in turbulence.
The amplitude of $\nabla\cdot\mathbf{U}_\psi$ is of order 0.1--1 $\Omega_{ce}$ at the three reconnection X-points.

\section{Discussion}
The flux transport velocity has been generally considered as the E$\times$B drift velocity. In \eqnref{eq:Upsi}, the slippage between magnetic flux and electron flow arising from an non-ideal electric field $\mathbf{E}_e'$ is included. For the three reconnection X-points, the slippage provides the major contribution to the inflows and outflows of magnetic flux near the X-point, being $\sim$2--3 times larger than the perpendicular electron flow. Further away from the X-point where the the slippage becomes small, $\mathbf{U}_\psi$ follows the perpendicular electron flow, which is mainly the E$\times$B drift. 

$\nabla\cdot\mathbf{U}_\psi$ consistently shows a quadrupolar structure at all reconnection X-points in turbulence. However, a signal is possible at O-points, where magnetic flux annihilation could happen. This process is recently explored by MMS \citep{Hasegawa:2020}. MFT activity at O-points deserves future investigation.


A new category of reconnection in turbulence beyond electron-only reconnection \citep{Phan:18} is revealed by X3. Only a single electron Alfv\'enic electron jet and Alfv\'enic ion jet are observed at X3. This category has $\mathbf{U}_\psi$ reversals, but no plasma outflow jet reversal. Electron-only reconnection with only one jet is also reported in simulations of shock-driven turbulence \citep{Bessho:2020}.

\section{Application to heliospheric plasmas}
Application of the MFT method to heliospheric plasmas requires the following conditions: (i) $k_\parallel \ll k_\perp$, where "$\parallel$" is along the background magnetic field, and (ii) the reconnection magnetic fields primarily reside on a local reconnection plane. $k_\parallel \ll k_\perp$ is based on $k_\parallel/k_\perp \ll \delta E_\parallel/\delta E_\perp$ for deriving $\partial_t\psi$ (\secref{sec:theory}), a condition well satisfied in the simulation.
\eqnref{eq:Upsi} is then a good approximation of $\mathbf{U}_\psi$ even in 3D systems. Physically, this represents quasi-planar reconnection with parallel length scales much longer than perpendicular. $k_\parallel \ll k_\perp$ is well satisfied in the cascade of kinetic Alfv\'en wave turbulence \citep{Cho:2004,Schekochihin:2009}, which is consistent with solar wind and magnetosheath observations \citep{Alexandrova:2008b,Alexandrova:2009,Sahraoui:2013b,Chen:2016,Chen:2017}. The model of planar reconnection is adopted by the local current sheet (LMN) coordinate \citep{Sonnerup:1967}, commonly used in space reconnection observations. Observations of reconnection in small-scale current sheets in the turbulent magnetosheath are consistent with this model (e.g. \citep{Phan:18,Wilder:2018}). Thus, the conditions for applying MFT is expected to be realistic for reconnection in heliospheric turbulence. Recent 3D PIC simulations further show that a long extended X-line, satisfying $k_\parallel \ll k_\perp$, easily arising in sub-ion-scale current sheets in 3D \citep{Li:2020a}, also favors reconnection activity \citep{YHLiu:19a,Huang:2020}.

\section{Conclusion}
The MFT method is a new way of identifying reconnection X-points in turbulent plasmas. It captures bi-directional inflows and outflows of magnetic flux at the X-points to signify reconnection, even without bi-directional plasma outflow jets. $\nabla\cdot\mathbf{U}_\psi$ is suitable for use in multi-spacecraft missions such as MMS. The first application to a 2D gyrokinetic turbulence simulation demonstrates the capability of this method in clearly capturing active reconnection signatures, as an inflow-outflow pattern or a quadrupolar structure in $\nabla\cdot\mathbf{U}_\psi$. It also reveals a new category of reconnection in turbulence beyond electron-only reconnection. This method has the potential to replace the plasma outflow jet reversal signature for reconnection. Applications to 3D simulations and heliospheric observations from spacecraft missions will present new opportunities to study the role of reconnection and identify new types of reconnection in turbulence.


The authors thank Tai Phan, Prayash Sharma Pyakurel and Daniel Verscharen for fruitful discussions. This work is supported by NSF award AGS-2000222 and NASA grants 80NSSC18K0754 and MMS mission 80NSSC18K0289. 


\providecommand{\noopsort}[1]{}\providecommand{\singleletter}[1]{#1}%

\appendix

Two supplementary figures are available. Figure A1 shows the turbulent cascade in the dissipation range in the magnetic energy spectrum. Figure A2 shows the fluctuating in-plane ion flow velocity $\delta \mathbf{u}_i$ for X3.

\begin{figure}
\figurenum{A1}
\resizebox{2.5in}{!}{\includegraphics[scale=1.,trim=0.2cm 0.1cm 0.1cm 0.1cm, clip=true]{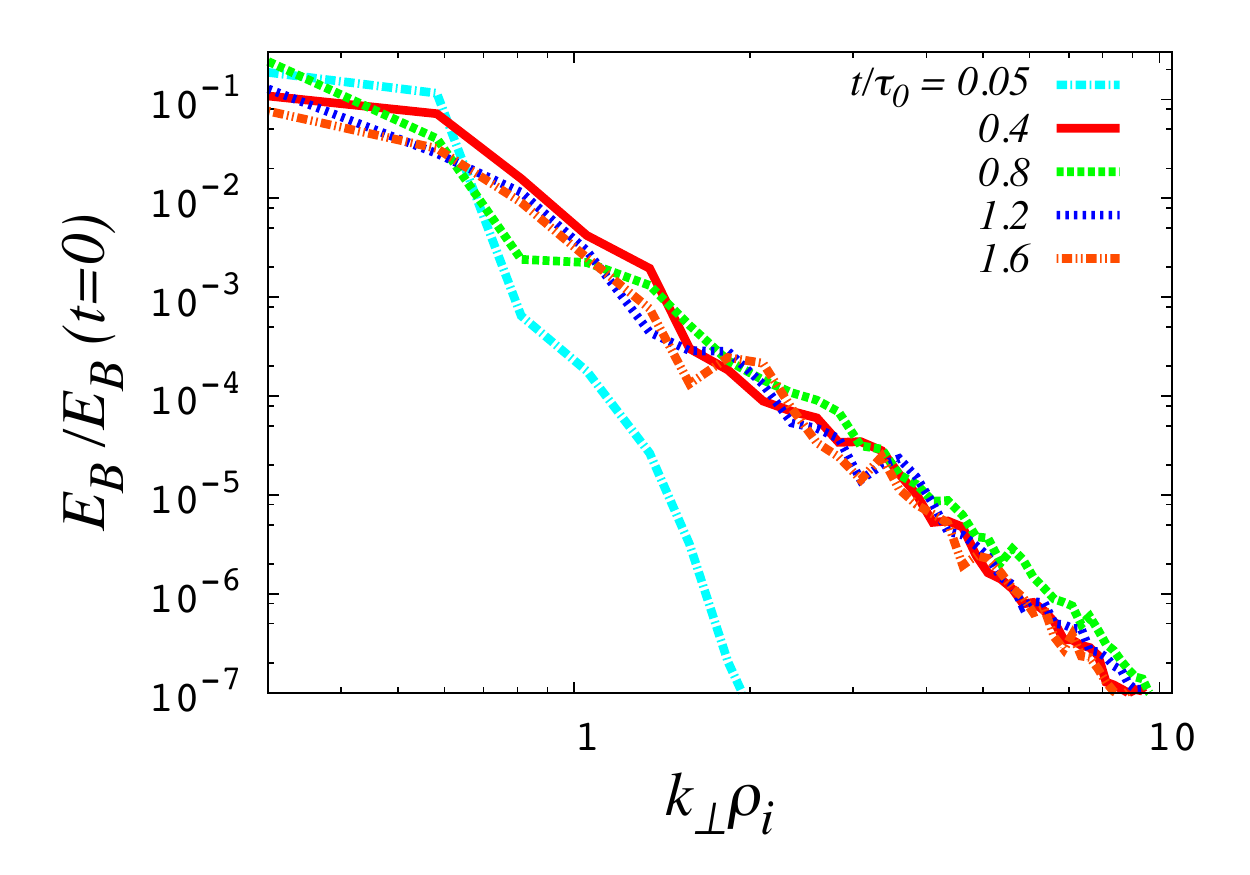}}
\caption{ \label{fig:fA1} Evolution of the magnetic energy spectrum as a function of scale, showing a developed turbulent cascade in the dissipation range.}
\end{figure}

\begin{figure}
\figurenum{A2}
\resizebox{2.5in}{!}{\includegraphics[scale=1.,trim=0.5cm 1.cm 0.1cm 2cm, clip=true]{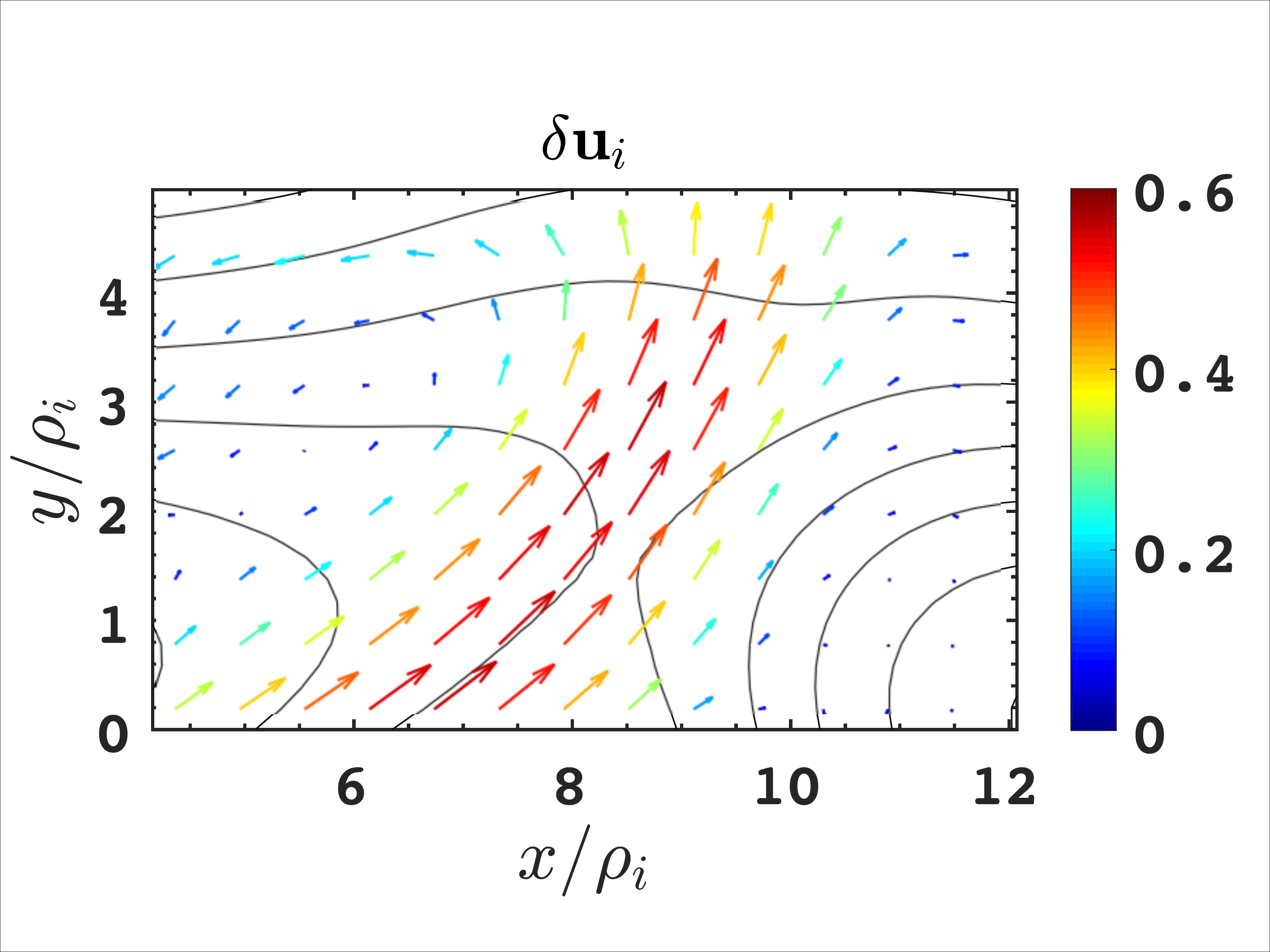}}
\caption{\label{fig:fA2}Ion flow velocity $\delta \mathbf{u}_i$ for X3, showing one outflow jet in ions. Compare with the electron flow velocity in \figref{fig:f4}(c). $\delta \mathbf{u}_i$ is normalized to the ion thermal speed $v_{ti}$. }
\end{figure}

\end{document}